# Title : Making the best of data derived from a daily practice in clinical legal medicine for research and practice – the example of Spe3dLab

Authors: Vincent Laugier, MSc (1,2), Eric Stindel MD PhD (3), Alcibiade Lichterowicz, MSc (1), Séverine Ansart, MD PhD (3) and Thomas Lefèvre MD PhD (4,5)

((1) Tekliko SARL - 362, chemin de la Bosque Antonelle, Quartier Celony, 13090 Aix-en-Provence, France, (2) Tekliko Pte Ltd - 100, Pasir Panjang Road, #07-07, Singapore, 118 518, (3) Laboratory of Medical Information Processing (LaTIM - INSERM UMR 1101), 29200 Brest, France, (4) Department of forensic medicine, Hôpital Jean Verdier APHP, Avenue du 14 Juillet, 93140, Bondy, France, (5) IRIS - Institut de Recherches Interdisciplinaires sur les enjeux Sociaux, INSERM, CNRS, EHESS, Université Paris 13, UMR 8156-997, 93100, Bobigny, France)

Corresponding author:
Mail: vincent.laugier@tekliko.com
Phone: +65 9119 7563

**Abstract**

Forensic science suffers from a lack of studies with high-quality design, such as randomized controlled trials (RCT). Evidence in forensic science may be of insufficient quality, which is a major concern. Results from RCT are criticized for providing artificial results that are not useful in real life and unfit for individualized prescription. Various sources of collected data (e.g. data collected in routine practice) could be exploited for distinct goals. Obstacles remain before such data can be practically accessed and used, including technical issues. We present an easy-to-use software dedicated to innovative data analyses for practitioners and researchers. We provide 2 examples in forensics. Spe3dLab has been developed by 3 French teams: a bioinformatics laboratory (LaTIM), a private partner (Tekliko) and a department of forensic medicine (Jean Verdier Hospital). It was designed to be open source, relying on documented and maintained libraries, query-oriented and capable of handling the entire data process from capture to export of best predictive models for their integration in information systems. Spe3dLab was used for 2 specific forensics applications: i) the search for multiple causal factors and ii) the best predictive model of the functional impairment (total incapacity to work, TIW) of assault survivors. 2,892 patients were included over a 6-month period. Time to evaluation was the only direct cause identified for TIW, and victim category was an indirect cause. The specificity and sensitivity of the predictive model were 99.9% and 90%, respectively. Spe3dLab is a quick and efficient tool for accessing observational, routinely collected data and performing innovative analyses. Analyses can be exported for validation and routine use by practitioners, e.g., for computer-aided evaluation of complex problems. It can provide a fully integrated solution for individualized medicine.

# Making the best of data derived from a daily practice in clinical legal medicine for research and practice – the example of Spe3dLab

## Abstract

Introduction

Forensic science suffers from a lack of studies with high-quality design, such as randomized controlled trials (RCT). RCT are difficult to set up for practical or ethical issues. Evidence in forensic science may be impacted or be of insufficient quality, which is a major concern. However, the results from RCT are criticized for providing artificial results that are not useful in real-life settings and are unfit for individualized diagnosis or prescription. Meanwhile, various sources of collected data, such as data collected in routine practice settings, could be exploited given distinct goals. Several obstacles remain before such data can be practically accessed and used, including technical issues. In this article, we present a software dedicated to easy-to-use and innovative data analyses for practitioners and researchers. We provide two case studies of this software in clinical legal medicine.

Methods

The Spe3dLab software has been jointly developed by three French teams: a bioinformatics laboratory (LaTIM, Inserm U1101), a private partner specialized in health software (Tekliko) and a department of forensic medicine (Jean Verdier Hospital, France). It was designed to be open source, relying on documented and maintained libraries, query-oriented, easy-to-use and capable of handling the entire data process from capture to export of best predictive models for their integration in information systems. Spe3dLab was used for two specific forensics applications: i) the search for multiple causal factors and ii) the best predictive model of the functional impairment (total incapacity to work, TIW) of assault survivors derived from data collected in a department of forensic medicine.

Results

Spe3dLab has been successfully installed and integrated in the workflow of the Jean Verdier department of forensic medicine. A total of 2,892 patients were included over a 6-month period. Time to evaluation was the only direct cause identified for TIW, and victim category was an indirect cause. The specificity and sensitivity of the predictive model were 99.9% and 90%, respectively.

Discussion

Spe3dLab is an interesting, quick and efficient tool for accessing observational, routinely collected data and performing innovative analyses. Analyses can be exported for prospective validation and routine used by practitioners, e.g., for computer-aided evaluation of complex problems. It can provide a fully integrated solution for appropriate individualized medicine.

## Introduction

The pitfalls of evidence-based medicine (EBM) were exposed as soon as it was theorized in 1996[1]. Its weaknesses can be summarized as both a problem inherent to the attempt of creating standardized guidelines supposed to be relevant to any situation as well as a reluctance of practitioners in adopting them in their daily practice.

Asking the local stakeholders to take part in the validation of the guidelines has proven to be successful in overcoming the acceptance issue[2],[3]. Nevertheless, assessing the relevance of a guideline to a particular context remains. In the context of clinical legal medicine, where we need to assess the impairment of victims of assaults or determine one patient's fitness for police custody, EBM is especially difficult to obtain due to the heterogeneity of practices by forensic examiners and sparsity of high level evidence. Proposals of guidelines in forensics are often limited to proposals of methodology[4].

Big data and, more generally, the access to the large amounts of electronic health records (EHR) that a health care organization produces have provided a way for personalized or individualized medicine to complement or substitute EBM with a 'data-driven' approach[5]. Instead of relying only on established findings that may prove irrelevant or incomplete, researchers and practitioners can now run analyses that take into account the specifics of the context and the patient. EHR in clinical legal medicine are minimally used in France. Most of the time, data are collected for a specific purpose in the context of a specific study. Proposals of standardization such as in[6] may lead to the routine and homogeneous production of large amounts of clinical legal medical data in the future.

The prerequisite to any EHR-based analysis is the access to the data. Software components such as i2b2[6] can help in gathering, merging and organizing data, but the researcher and the practitioner are eventually left with the tedious task of analyzing their data. This represents a serious impediment to wider implementation of individualized medicine and is subsequently a great loss for clinical forensic research because these experts have an easy access to the data and are the best fit to formulate relevant hypotheses.

Projects such as[7] simplify this analysis by providing a web interface for a given type of analysis.

Easy-to-use software should at least provide a way for researchers and practitioners to answer the following three questions posed by data analysis: "What is the description of the considered population?" "What are the (causal) relationships between the variables?", and "What prediction can I make based on observations?".

We have designed a high-level process shared by all types of data analysis encompassed by the three aforementioned questions (Table 1). This model has been implemented in a software layer called Spe3dLab that can plug into several types of data sources.

In this article, Spe3dLab was used for two specific forensic applications: i) the search for multiple causal factors and ii) the best predictive model of the functional impairment (total incapacity to work, TIW) of assault survivors derived from data collected in a department of forensic medicine.

Table 1
The different types of analysis proposed by Spe3dLab (as of May 2016)

| Types of questions asked | Types of analysis |
| --- | --- |
| What is the description of the considered population? | • For continuous variables: Min, Max, Mean, Median, Quartiles, Standard deviation<br>• For categorical variables: Frequency of every category |
| What are the (causal) relationships between the variables? | • Clustering<br>• Hypothesis testing<br>• Bayesian networks (causal) |
| What prediction can I make based on observations? | • Prediction of outcome values for a continuous variable<br>• Classification for a categorical variable |

Spe3dLab is a query-oriented software. The table shows the main correspondences between the type of questions one could ask and the type of analysis meant to answer the question.

**Methods**

Spe3dLab software development

Spe3dLab[8] was originally initiated by a partnership between a French bioinformatics laboratory (LaTIM, Inserm U1101) and a company specializing in software development for healthcare in 2013 (Tekliko). As the project gathered momentum, a third public partner joined in. To date, Spe3dLab is the evolving result of the LaTIM, Tekliko and the department of forensic medicine of the Jean Verdier Hospital (AP-HP, France).
Spe3dLab was designed to be open source, relying on documented and maintained libraries, query-oriented, easy-to-use and capable of handling the whole data process from capture to export of best predictive models for their integration in information systems.

Software components

Depending on the data source of interest, the data are stored as structured data in PostgreSQL databases, CSV (comma separated values) files or in-memory datasets shipped with R libraries. The underlying data-analysis software is R[9]. The web-based graphical user interface is built with R Shiny[10], the API of the predictive models built with Spe3dLab is running in a Wildfly JEE Server that interacts with Rserve[11]. The overall architecture is given in Figure 1. The workflow allowed by Spe3dLab is depicted as an activity diagram in Figure 2.
Different R libraries are used here, dplyr[12] for pretreatment of the data, ggplot2[13] and ggrepel[14] for the plotting of the various graphs, e1071[15] for Support Vector Machine algorithms and the SuperLearner[16] for predictive analysis, FactoMineR[17] for Principal Component Analysis (PCA) and Multiple Correspondence Analysis (MCA), bnlearn[18] for Bayesian Network analysis, cluster[19] for clustering analysis, and stats[20] (included in R core libraries) for ANOVA and Student's t-test.

Figure 1
Simplified architecture of Spe3dLab

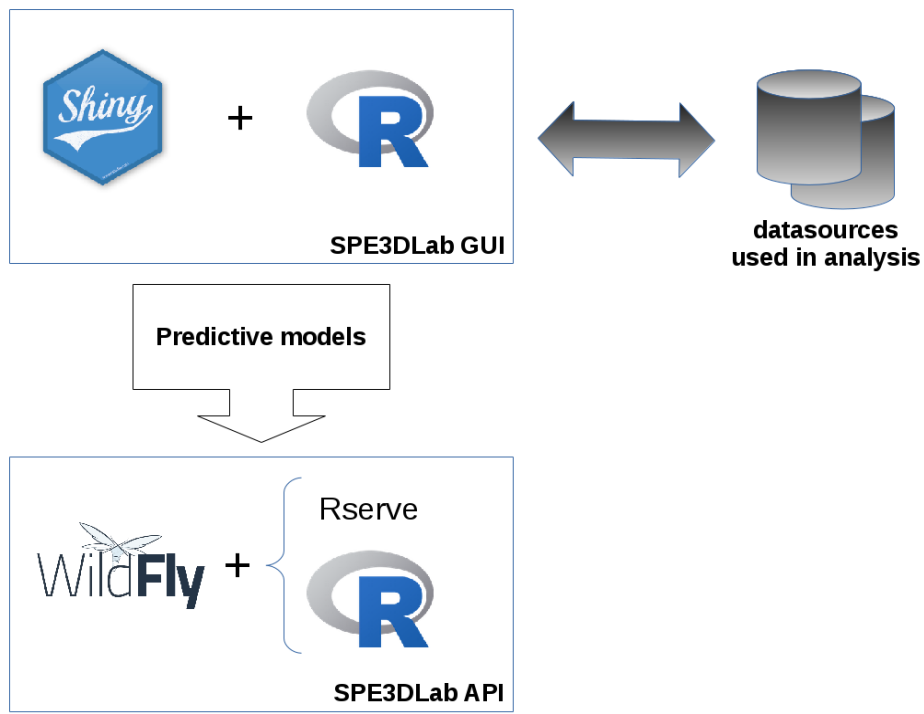

Modelers use Spe3dLab GUI to create predictive models, and these models are made available to third party application via Spe3dLab API. Shiny is an open source web application framework for R; WildFly is an open source Java web server; Rserve is a TCP/IP server that allows programs to use facilities of R. API: Application programming interface. GUI: Graphical user interface.

Figure 2
Activity diagram of a data analysis workflow based on Spe3dLab

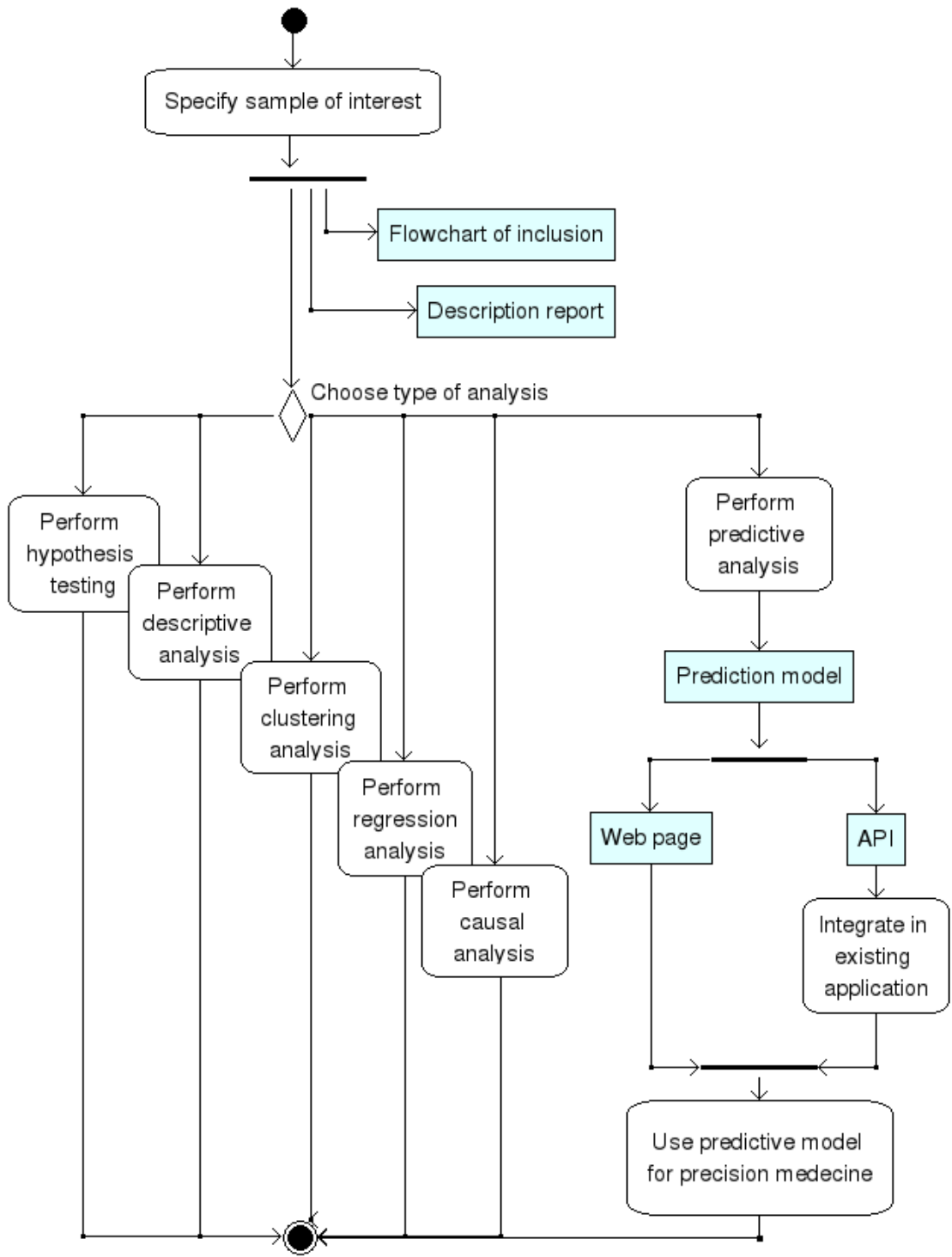

This figure depicts the different workflows possible from the specification of the sample to be included (included population according to inclusion/exclusion criteria) to the results of a given type of analysis (see Table 1 for types of user questions and types of analysis for answering these questions). In the specific case of prediction modeling (e.g., TIW predicting), the model obtained at the end of the process can be externalized from the Spe3dLab software, either as a web page presenting all relevant fields and variables to be filled or an API that can be used in a third-party software.

Description of the analysis workflow in Spe3dLab

- Data source selection: the user can directly access routinely collected data if Spe3dLab has been previously integrated with the information system. Otherwise, the user can access specific databases or upload his or her own data (Microsoft xlsx or CSV files).
- Analysis selection: Spe3dLab guides the user in the following basic analysis: descriptive analysis, hypothesis testing, clustering and search for specific patients' profiles, causal and quasi-causal analysis and predictive analysis. Spe3dLab provides a best model approach, which means that it will provide the user with the best possible model derived from available data under specific and clear assumptions. Best models are obtained through k-fold cross validations and resampling schemes. For causal and predictive models, personalized calculations can be performed. The user is invited to input data specific to the patient he/she is interested in. Causal analysis will provide the probability associated with a specific event given the values of all relevant causal factors. Predictive analysis will provide the value of the outcome according to the values of all parameters constitutive of the model.
- Variable selection: depending of the type of analysis, the user is invited to specify i) the outcome of interest, ii) inclusion/exclusion criteria and iii) variables to screen for tests or modeling. According to the specified inclusion/exclusion criteria, a flow chart is automatically built and comparison tests are run between populations with and without missing data. The selected type of analysis is then performed.
- Post analysis: once a model is obtained that meets the user's expectation, it can be exported and made autonomous from the data so that the model can be used as a standalone product. A web page that can be automatically built or an API (Application Programming Interface) is automatically generated so that the model can be further integrated in a third-party application, e.g., in an information system.

Two case studies in clinical legal medicine: causal and predictive models of functional impairment in assault survivors.

- Study population: The study population consists of assault survivors who have received a forensic medical examination to evaluate their functional impairment, quantified in days of total incapacity to work as described by French law[21]. All patients were examined by forensic physicians in one unique reference center (department of forensic medicine of the Jean Verdier Hospital, Bondy, France).
- Inclusion criteria were: being an assault survivor, being older than 10 years at the time of examination, having been examined by a forensic physician of the department of forensic medicine of the Jean Verdier Hospital, having been examined within the first 30 days from the assault, and having been examined between the 1st of January 2015 and the 30th of June 2015.
- Exclusion criteria: insufficient comprehension of French or absence of a professional translator, presenting an important psychiatric disorder that prevents the patient from properly relating the facts, patients with bone fracture, being a sexual assault survivor, and having suffered from involuntary violence. All inclusion and exclusion criteria were chosen in accordance with[22]. Analyses were performed on complete data (i.e., no missing data). Aberrant values were also dismissed, e.g., patients older than 120 years or data coded as '999' standing for non-available data.

Table 2
Description of the studied population (n = 2,892)

|  |  | N (%) |
|---|---|---|
| Gender | Men | 1750 (60.5) |
| Forensic examiner | *16 categories* |  |
| Assault place | Public way | 1133 (39.2) |
|  | Other (*15 categories*) | 793 (27.3) |
|  | Marital home | 309 (10.7) |
|  | Victim home | 271 (9.4) |
|  | Family home | 193 (6.7) |
|  | Workplace | 193 (6.7) |
| Assailant category | Unknown person | 949 (32.8) |
|  | (ex)Spouse, (ex)partner | 566 (19.6) |
|  | Police officer | 411 (14.2) |
|  | Person known from sight | 214 (7.4) |
|  | Other known person | 179 (6.2) |
|  | Other (*15 categories*) | 573 (19.8) |
| Assault category | Other (*10 categories*) | 1339 (46.3) |
|  | Marital violence | 576 (19.9) |
|  | Police violence | 419 (14.5) |
|  | Gang assault | 365 (12.6) |
|  | Family violence | 193 (6.7) |
| Injury type | Presence of injury | 2274 (78.6) |
| Victim category | Other | 1978 (68.4) |
|  | Person in Custody | 798 (27.6) |
|  | Police officer | 116 (4) |
| Age (years) | [11-22] | 758 (26.2) |
|  | [23-30] | 716 (24.8) |
|  | [31-40] | 704 (24.3) |
|  | ≥ 41 | 714 (24.7) |
| Time to evaluation (hours) | [0-11] | 737 (25.5) |
|  | [12-47] | 673 (23.3) |
|  | [48-71] | 606 (21) |
|  | ≥ 72 | 876 (30.3) |
| Number of aggravating factors | 0 | 509 (17.6) |
|  | 1 | 891 (30.8) |
|  | 2 | 826 (28.6) |
|  | 3 or more | 666 (23) |
| Total incapacity to work (days) | [0-8] | 2731 (94.4) |
|  | ≥ 9 | 161 (5.6) |

N: number of patients. Variables 'Age', 'Time to evaluation', 'Number of aggravating factors', 'TIW' have been transformed into categorical variables. The cutoff points used for 'Age' and 'Time to evaluation' are quartiles.

- Variables and patients' characteristics
Consistent with[22], variables considered for analysis were the following:
Patients' characteristics: age (years), gender (male/female), victim category (police officer, individual in police custody, other), type of injury (none, hematoma or bruise, wounds).
Circumstances of the assault: assault category (14 categories), number of aggravating factors (in the sense of the French Penal Code, 0-1-2-3 or more than 3 factors), assailant category (20 categories), assault place (20 categories).
Characteristics of the medical examination: forensic physician (16 distinct physicians), time to evaluation (in hours) and functional impairment quantified in days of total incapacity to work.
For comparison purposes with the results from[22], all continuous variables (age, TIW, time to evaluation) were automatically categorized as shown in Table 2.

Case studies in clinical legal medicine

The functional impairment of assault survivors can be quantified in France by the number of days of 'Total Incapacity to Work' ('Incapacité totale de travail,' ITT). French law uses the threshold of 8 days of TIW to classify the "seriousness" of the offense [21].
We first try to identify the causal relationships between the TIW and the characteristics of the victim, the assault, the assailant and the circumstances of the evaluation. We then try to find a predictive model capable of predicting whether the TIW will be greater than 8 days or not. This case study of exploratory data-analysis is inspired by[22].
- First case study: Searching for causal factors that account for the functional impairment in assault survivors.
Bayesian networks were used to search for causal factors that explain functional impairment in assault survivors. A first study was led exposing the principles of Bayesian networks as a data mining method[22]. A Bayesian network is a probabilistic acyclic graph composed of nodes and arcs representing the variables and the relationships of causalities between the variables, respectively, and by a joint probability distribution (JPD) of these variables[23,24]. The inferred network and parameters of the network can then be used to calculate conditional probabilities. We used a score-based algorithm named Hill-Climbing algorithm[25] for network inference. The corresponding best model approach is ensured by the exploration of the DAG space, retaining the best scoring DAG among others.
A major strength of Bayesian networks is that if no latent variable exists that has not been included among the considered variables, then relationships between variables are causal relationships in a probabilistic sense, not deterministic. The duration of TIW was taken as the variable of interest, and all other variables were included as potential causes of the duration of TIW.
- Second case study: Predicting whether the TIW is above or below the threshold for judicial qualification of offense.
We used a well-known but under-exploited technique to build our predictive model. Support vector machines (SVM) were used to predict whether the TIW, according to patients' characteristics, circumstances of the assault and the examination characteristics, is below or above the threshold of 8 days. SVM refers to machine learning techniques and algorithms used for the prediction of group membership. SVM use a training dataset to build a model that is then used to classify new recordings[26,27]. Cost and gamma are two parameters of SVM models that control how the groups are separated by changing the number of recordings taken into account when drawing the boundaries between the groups as well as their weights. The best model approach is implemented here as the parameter space (cost/gamma space) is screened for the best combination. The best combination is the values for the cost/gamma parameters that maximize the sensibility and the specificity of the predictive model.

## Results

After consideration of the inclusion and exclusion criteria, 2,892 patients out of 4,279 were included in the study over a 6-month period. Flow chart of inclusion automatically generated by Spe3dLab and descriptive results of the included population are given in Figure 3 and Table 2. The median duration of TIW was 3 days [0-96].
Causal analysis of TIW. The resulting Bayesian network is shown in Figure 4. Causal analysis found that the time to evaluation was the only direct causal factor of TIW. The secondary cause of TIW ("causes of causes") was the victim category.
Predictive analysis of TIW according the legal threshold of 8 days. Sensitivity and specificity are given for each combination of cost/gamma values. Figure 5 shows that the best values for the gamma/cost was 1/10. The selected model had a sensitivity of 90% and a specificity of 99.9%. The confusion matrix is given in Table 3. Only 2 cases (<0.01% of all cases) were misclassified as greater than 9 days, and 16 (<0.1% of all cases) were misclassified as less than 9 days.

Table 3
Comparison of predicted values against real values of Total Incapacity to Work (TIW) for the best SVM model

| TIW (days) | Real values, [0-8] | Real values, ≥ 9 |
|---|---|---|
| Predicted values, [0-8] | 2729 | 16 |
| Predicted values, ≥ 9 | 2 | 145 |

The confusion matrix depicts the true and false positives (respectively, negatives) resulting from the prediction algorithm applied to TIW classification between 0 to 8 days and 9 or more days of TIW. Positive is defined here as a TIW equal or greater than 9 days. The table shows that there were only 2 false positives and 16 false negatives.

Figure 3
Flowchart of inclusion of the data analysis of functional impairment of assault survivors

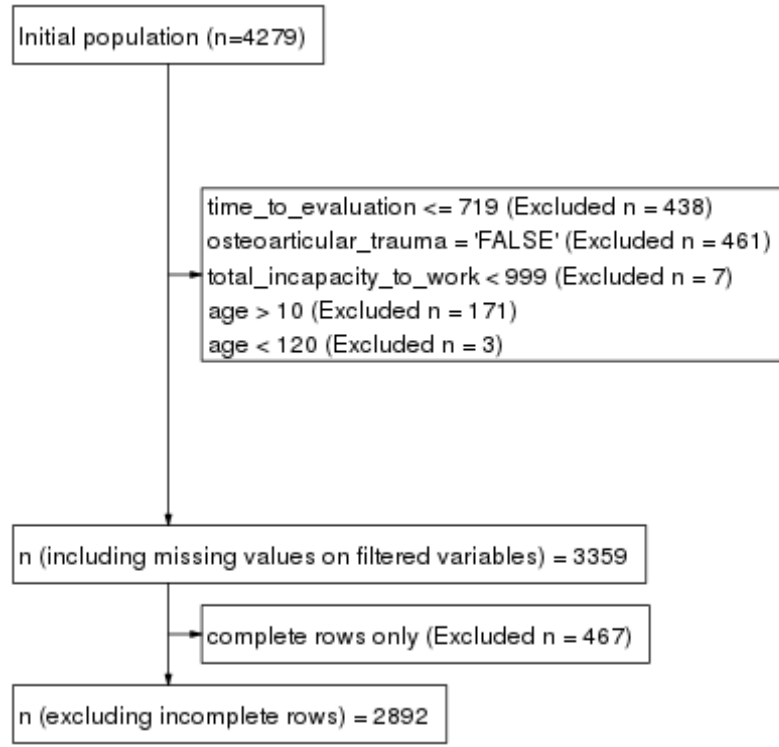

The figure shows the flowchart of inclusion as automatically generated by Spe3dLab based on inclusion/exclusion criteria. Here, only complete data are considered, which led to the exclusion of 467 patients.

Figure 4
The directed acyclic graphs associated with the data related to the functional impairment of assault survivors (n=2,892, Hill Climbing algorithm)

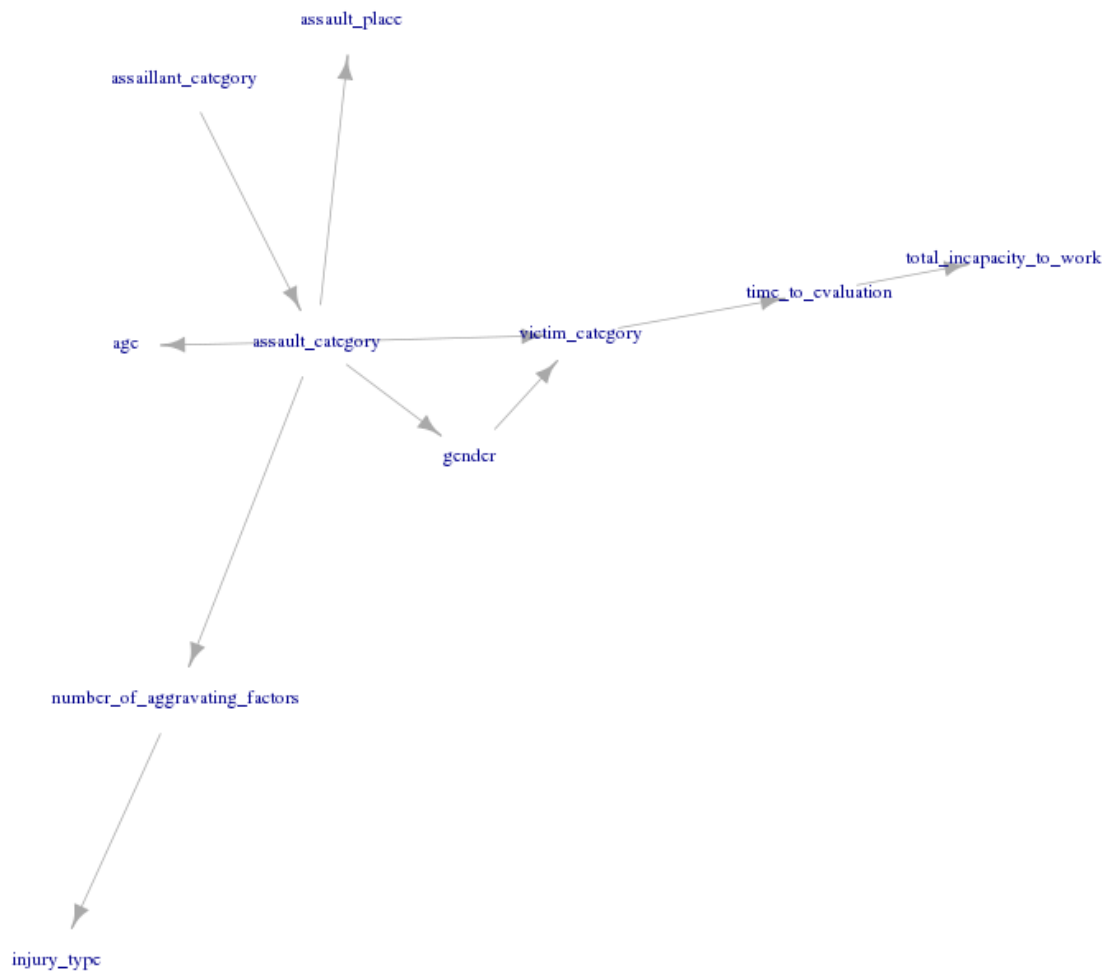

The figure shows the best Bayesian network obtained by the Hill-Climbing algorithm, as automatically generated by Spe3dLab. Time to evaluation is directly causes TIW, while the victim category is a direct cause of the time to evaluation. Interestingly, we can see a cluster of characteristics centered on "assault category". The assault category variable is directly linked to the assailant category, the assault place and the number of aggravating factors.

Figure 5
Plot of the sensitivity and specificity of the SVM predictive models for the different values of gamma/cost

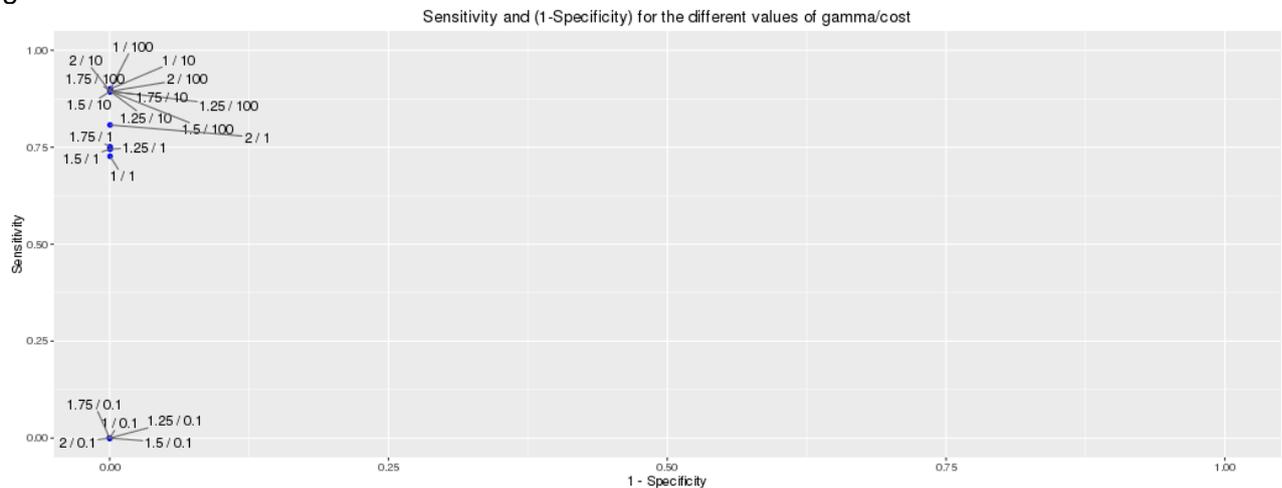

The figure shows the different values for sensitivity and specificity of the different tested predictive models. Of note, all models presented almost perfect specificity. Improvements could only be made in terms of sensitivity. As in any other ROC curve, the best models are located in the upper left corner. x/y values reported on the figure are the corresponding values of the of parameters Gamma/Cost. One of the best equivalent models was given for the ratio 1/10.

## Discussion

The lack of exploitation of routinely collected clinical forensic data is a loss for practitioners and researchers as well as the different stakeholders working with them (police, emergency units, decision makers). These data could at least be used for retrospective analysis or group identification with the intent of setting up a new study[28]. Practitioners could gain visibility on their activity and benefit from an analysis tool. Spe3dLab allows practitioners and researchers to easily answer their descriptive and predictive questions using classical and non-classical techniques of data-analysis.

Spe3dLab has proven effective at easily reproducing the method of[22] while being much easier to access. Differences exist between the two studies in terms of results. There are several possible explanations. While the data were collected in the same unit, the data analyzed in the present article were recorded over different time periods. The Bayesian network was built using a score-based algorithm named the Hill Climbing algorithm instead of the IAMB, which is a constraint-based algorithm. This choice was made in order to generate a directed acyclic graph (DAG) that clearly shows causality relationships between the variables[29]. Here, time to examination was the only direct cause identified for TIW. It is notable that the victim category was found to be an indirect cause for TIW, causing the time to examination. Time to examination appeared here as a mediating factor between TIW and the victim category. The analysis conducted here relied on a larger sample (about three times larger) and may be more valid than previous results published. Moreover, previous results regarding TIW and Bayesian networks were primarily an illustration for exposing the principles of Bayesian networks. A more comprehensive study encompassing more complete and balanced characteristics about the assault survivor, the circumstances of the assault, the characteristics of the examination and various psychological and physical aspects of the functional impairment, is still needed. Ultimately, an even more robust approach should be implemented based on merged, consensus and average graphs, which is still to be implemented in Spe3dLab.

To date, novel predictive methods such as SVM have been very rarely if ever used in forensics[30]. This may be mainly due to a lack of knowledge from forensic scientists and to an unease access to these methods. Here, we showed that such techniques can be highly efficient in predicting whether or not the duration of the TIW may be over the judicial threshold given a number of characteristics.

Limitations

For the sake of study comparison and reproducibility, we used categorical variables in the causal analysis. This approach was not mandatory because the Hill-Climbing algorithm can handle mixed variables, i.e., categorical and continuous variables, at the same time. Interestingly, the results obtained with mixed variables were partly consistent with those presented here, but partly not. The performance of the predictive model was very good. Nonetheless, a more practical aspect of TIW prediction would be an accurate prediction of the values of the TIW duration and not only with regard to a threshold. Predicting a TIW values and providing the corresponding confidence interval would help achieve double contextualization or individualization. Taking into account both the characteristics of the physicians and of the assault survivors would provide more homogenous and observation-based results, adapted to both the examiner and the victim. However, any predictive model has to be validated on prospective data. Indeed, machine learning techniques are known to be prone to overfitting; their performances are very high, but an overly specific data set was used to train the algorithm, and so it can lack generalizability when applied to newly collected data. This phenomenon is known as overfitting. Real performance is expected to be slightly lower than that presented here. Validation is facilitated by the integration capability offered by Spe3dLab: the predictive model can be integrated in the information system and tested in real-life every day. The result rendering in this study, i.e., the layouts used for Bayesian networks or flow charts for the tables for statistics presentation, were chosen to provide the most results produced by the solution without any changes so that one may have a fair idea of the current possibilities of Spe3dLab in rendering. As of today, the rendering is still insufficient and will be improved over time.

Perspectives

In the next months, we foresee full integration of Spe3dLab with a forensic open source information system[31] so that we approach real personalized or contextualized medicine. Spe3dLab will make use of an open data model for clinical legal medicine. Other analytics modules will be added, such as time series models and systems dynamics, automated information retrieval in full text sources such as PubMed and social networks e.g., Twitter[32]. Tests stressing data quality and completeness will be conducted between two distinct approaches: analyses based on structured versus unstructured data. These tests will be enabled via our new information system and the use of text mining techniques currently integrated in the core functions of Spe3dLab. The use of very heterogeneous data sources, such as those coming from different forensics departments across several countries, will not be possible without any practical consideration for a proper semantic interoperability framework[33]. Spe3dLab is based on open-source projects and libraries; its source code is open as well (under GPL) and will be made available soon.

**References**

1. Timmermans S., Mauck A. The promises and pitfalls of evidence-based medicine. *Health Aff* 2005:18–28. Doi: 10.1377/hlthaff.24.1.18.

2. Powell C V., Maskell GR., Marks MK., South M., Robertson CF. Successful implementation of spacer treatment guideline for acute asthma. *Arch Dis Child* 2001;**84**(2):142–6. Doi: 10.1136/adc.84.2.142.

3. Gross PA., Greenfield S., Cretin S., Ferguson J., Grimshaw J., Grol R., et al. Optimal methods for guideline implementation: conclusions from Leeds Castle meeting. *Med Care* 2001;**39**(8 Suppl 2):II85–92. Doi: 10.1097/00005650-200108002-00006.

4. Keune LH., de Vogel V., van Marle HJC. Methodological development of the Hoeven Outcome Monitor (HOM): A first step towards a more evidence based medicine in forensic mental health. *Int J Law Psychiatry* 2016;**45**:43–51. Doi: 10.1016/j.ijlp.2016.02.009.

5. Raghupathi W., Raghupathi V. Big data analytics in healthcare: promise and potential. *Heal Inf Sci Syst* 2014:3. Doi: 10.1186/2047-2501-2-3.

6. I2B2 website. Available from: https://www.i2b2.org/. Accessed April 1, 2016.

7. Martinez-Millana A., Fernandez-Llatas C., Sacchi L., Segagni D., Guillen S., Bellazzi R., et al. From data to the decision: A software architecture to integrate predictive modelling in clinical settings. *Proceedings of the Annual International Conference of the IEEE Engineering in Medicine and Biology Society, EMBS*, vol. 2015-Novem. 2015. p. 8161–4.

8. Spe3dLab website. Available from: www.speelab.io. Accessed November 1, 2016.

9. R Project website. Available from: https://www.r-project.org/. Accessed June 1, 2016.

10. Shiny R framework website. Available from: http://shiny.rstudio.com/. Accessed June 1, 2016.

11. Rserve website. Available from: https://rforge.net/Rserve/. Accessed November 22, 2016.

12. dplyr R package web page. Available from: https://cran.r-project.org/web/packages/dplyr/index.html. Accessed June 20, 2006.

13. Wickham MH. Package "ggplot2" 2014.

14. ggrepel R package web page. Available from: https://cran.r-project.org/web/packages/ggrepel/index.html. Accessed June 1, 2016.

15. Hornik K., Weingessel A., Leisch F., Davidmeyerr-projectorg MDM. *Package "e1071."* 2015.

16. Polley E., LeDell E., van der Laan M. Package "SuperLearner": Super Learner Prediction 2016:32.

17. Husson F., Josse J., Le S., Mazet J. Package " FactoMineR ." *R Top Doc* 2015:1–95. Doi: 10.1007/978-3-540-74686-7.

18. Package T., Learning P., Scutari AM. Package " bnlearn " 2016.

19. Maechler M., Struyf A., Hubert M., Hornik K., Studer M., Roudier P. Package " cluster ." *R Top Doc* 2015:79. Doi: ISBN 0-387-95457-0.

20. R Stats Package web page. Available from: https://stat.ethz.ch/R-manual/R-devel/library/stats/html/00Index.html. Accessed June 1, 2016.

21. French Penal Code - Article 222-13. Available from: https://www.legifrance.gouv.fr/affichCodeArticle.do?cidTexte=LEGITEXT000006070719&idArticle=LEGIARTI000006417637.

22. Lefèvre T., Lepresle A., Chariot P. Detangling complex relationships in forensic data: principles and use of causal networks and their application to clinical forensic science. *Int J Legal Med* 2015;**129**(5):1163–72. Doi: 10.1007/s00414-015-1164-8.

23. Alston CL., Mengersen KL., Pettitt AN. *Case Studies in Bayesian Statistical Modelling and Analysis*. Wiley; 2012.

24. Taroni, F., Aitken, C., Garbolino, P., B iedermann a. *Bayesian Networks and Probabilistic Inference in Forensic Science Statistics in Practice*. 2006.

25. Gámez JA., Mateo JL., Puerta JM. Learning Bayesian Networks by Hill Climbing: Efficient Methods Based on Progressive Restriction of the Neighborhood. *Data Min Knowl Discov* 2011;**22**(1-2):106–48. Doi: 10.1007/s10618-010-0178-6.

26. Chiu YW. *Machine Learning with R Cookbook*. Packt Publishing; 2015.

27. Theodoridis S. *Machine Learning: A Bayesian and Optimization Perspective*. Elsevier Science; 2015.

28. Fiore LD., Lavori PW. Integrating Randomized Comparative Effectiveness Research with Patient Care. *N Engl J Med* 2016;**374**(22):2152–8. Doi: 10.1056/NEJMra1510057.

29. Scutari M. Learning Bayesian Networks with the bnlearn R Package. *J Stat Softw* 2010;**35**(3):1–22. Doi: 10.18637/jss.v035.i03.

30. Santos F., Guyomarc'h P., Bruzek J. Statistical sex determination from craniometrics: Comparison of linear discriminant analysis, logistic regression, and support vector machines. Forensic Sci Int 2014;245:204.e1–204.e8. Doi: 10.1016/j.forsciint.2014.10.010.

31. Dang C, Phuong T, Beddag M, Vega A, Denis C. A data model for clinical legal medicine practice and the development of a dedicated software for both practitioners and researchers. J Forensic Med Leg. DOI: http://dx.doi.org/10.1016/j.jflm.2016.11.002

32. Liu S, Young SD. A survey of social media data analysis for physical activity surveillance. J Forensic Med Leg. DOI: http://dx.doi.org/10.1016/j.jflm.2016.10.019

33. Jaulent MC, Leprovost D, Charlet J, Choquet R. Semantic interoperability challenges to process large amount of data perspectives in forensic and legal medicine. J Forensic Med Leg. DOI: http://dx.doi.org/10.1016/j.jflm.2016.10.002